\documentclass[%
reprint,
superscriptaddress,
amsmath, amssymb, aps,
prb,
]{revtex4-1}
\usepackage{url}
\usepackage{color}
\usepackage{graphicx}% Include figure files
\usepackage{dcolumn}% Align table columns on decimal point
\usepackage{bm}% bold math
\usepackage{array}
\newcolumntype{P}[1]{>{\centering\arraybackslash}p{#1}}

\usepackage{amsmath} % or simply amstext
\newcommand{\angstrom}{\textup{\AA}}

\newcommand{\beginsupplement}{%
        \setcounter{table}{0}
        \renewcommand{\thetable}{S\arabic{table}}%
        \setcounter{figure}{0}
        \renewcommand{\thefigure}{S\arabic{figure}}%
     }

\begin{document}

\preprint{APS/123-QED}

\title{Identification of ferrimagnetic orbitals preventing spinel degradation by charge ordering in Li$_x$Mn$_2$O$_4$}

\author{Hasnain Hafiz}
\email{hafiz.h@husky.neu.edu}
\affiliation{Department of Physics, Northeastern University, Boston, MA}
\affiliation{Current Address: Department of Mechanical Engineering, Carnegie Mellon University, Pittsburgh, Pennsylvania, 15213, USA}

\author{Kosuke Suzuki}
\affiliation{Faculty of Science and Technology, Gunma University, Kiryu, Gunma 376-8515, Japan}

\author{Bernardo Barbiellini}
\affiliation{School of Engineering Science, LUT University, FI-53851 Lappeenranta, Finland}
\affiliation{Department of Physics, Northeastern University, Boston, MA}

\author{Yuki Orikasa}
\affiliation{Department of Applied Chemistry, Ritsumeikan University, Kusatsu, Shiga 525-8577, Japan}

\author{Stanislaw Kaprzyk}
\thanks{Deceased, October 2018}
\affiliation{Faculty of Physics and Applied Computer Science, AGH University of Science and Technology, aleja Mickiewicza 30, Krakow 30-059, Poland}
\affiliation{Department of Physics, Northeastern University, Boston, MA}

\author{Naruki Tsuji}
\affiliation{Japan Synchrotron Radiation Research Institute, SPring-8, Sayo, Hyogo 679-5198, Japan}

\author{Kentaro Yamamoto}
\affiliation{Graduate School of Human and Environmental Studies, Kyoto University, Sakyo-ku, Kyoto 606-8501, Japan}

\author{Ayumu Terasaka}
\affiliation{Faculty of Science and Technology, Gunma University, Kiryu, Gunma 376-8515, Japan}

\author{Kazushi Hoshi}
\affiliation{Faculty of Science and Technology, Gunma University, Kiryu, Gunma 376-8515, Japan}

\author{Yoshiharu Uchimoto}
\affiliation{Graduate School of Human and Environmental Studies, Kyoto University, Sakyo-ku, Kyoto 606-8501, Japan}

\author{Yoshiharu Sakurai}
\affiliation{Japan Synchrotron Radiation Research Institute, SPring-8, Sayo, Hyogo 679-5198, Japan}

\author{Hiroshi Sakurai}
\affiliation{Faculty of Science and Technology, Gunma University, Kiryu, Gunma 376-8515, Japan}

\author{Arun Bansil}
\email{ar.bansil@northeastern.edu}
\affiliation{Department of Physics, Northeastern University, Boston, MA}

\date{\today}

\begin{abstract}
Spinel Li$_x$Mn$_2$O$_4$ is a key cathode material that is used extensively in commercial Li-ion batteries. A challenge with this material has been that the capacity of the battery fades with cycling, an effect that can be traced to the presence of an anti-ferromagnetic insulator phase in the fully lithiated LiMn$_2$O$_4$ (LMO) and the associated charge disproportionation that drives distortions of the MnO$_6$ octahedra. Here, by combining x-ray magnetic Compton scattering experiments with parallel first-principles computations, we show that the anti-ferromagnetic phase of LMO is surrounded by a robust ferrimagnetic metallic phase, which becomes stable when even a small amount of Li is removed from or added to the charge-ordered LMO. In this surprising ferrimagnetic state, charge-ordering and octahedral distortions are found to be strongly suppressed. We identify the nature of the ferrimagnetic orbitals involved through theoretical and experimental analyses of the magnetic Compton scattering spectra.
\end{abstract}

\pacs{Valid PACS appear here}
\maketitle

\section{Introduction}
Spinel Li$_x$Mn$_2$O$_4$ is an attractive cathode material for rechargeable batteries\cite{Goodenough,Islam} because it is less expensive and environmentally more friendly than lithiated cobalt and nickel oxides. Unfortunately, the lithiated ($x=1$) compound LiMn$_2$O$_4$ (LMO) suffers from the problem of capacity fading due to a structural phase transition\cite{sanjeev,wanli}. Above the room temperature, LMO assumes the cubic structure of the normal spinel in which lithium ions occupy tetrahedral positions, with the manganese ions located only at the octahedral positions. A decrease in the temperature to around $283$ K produces the famous metal-insulator (Verwey) transition\cite{Hervieu}, which is associated with the distortion of the MnO$_6$ octahedra. This structural transformation limits the application of LMO as a cathode material\cite{Ragavendran}. The Verwey transition also drives the onset of long-range antiferromagnetic (AFM) order below the Curie temperature of $65$ K\cite{Tomeno}, which is also supported by first-principles calculations\cite{liu2017}. Geometric frustration of the AFM order in the spinel structure\cite{Willis} leads to complex potential energy landscapes that exhibit multiple magnetic phase transitions\cite{zhang2018} whose nature remains unclear.

In this study, we explore magnetic properties of spinel Li$_x$Mn$_2$O$_4$ on a first-principles basis by exploiting recent advances in constructing exchange-correlation functionals. Specifically, we employ the strongly-constrained-and-appropriately-normed (SCAN) functional\cite{SCAN}, which has proven especially accurate for investigating the electronic, geometric and magnetic structures of elemental manganese\cite{SCAN_Aki}, MnO$_2$ polymorphs\cite{SCAN_MnO2}, 3$d$ perovskite oxides\cite{SCAN_Zunger}, oxides superconductors\cite{SCAN_LSCO,chris_scan, SCAN_Yubo}, and layered lithium intercalated transition metal oxides \cite{Major2018}. Parallel magnetic Compton scattering experiments on LMO samples are also reported over a wide range of Li concentrations varying from unlithiated ($x=0$) to over-lithiated ($x=1.079$) Li$_x$Mn$_2$O$_4$.  Our analysis reveals that ferrimagnetism competes with the AFM order in LMO and leads to ferrimagnetic (FIM) moments even for slight departures from stoichiometry ($x=1$). Liu {\em et al.}\cite{liu2017} report that the AFM ordering is responsible for triggering changes in the Mn valence and driving Jahn-Teller distortions. The FIM phase we have found here, however, suppresses the octahedral distortions, which are responsible for cathode degradation. We show how the magnetic state associated with this puzzling FIM phase can be visualized through an analysis of our magnetic Compton spectra. 

Compton scattering, which refers to inelastic x-ray scattering in the deeply inelastic regime, provides a direct probe of the ground-state momentum density $\rho({\mathbf p})$ of the many-body electronic system through a measurement of the so-called Compton profile, $J(p_z)$, where $p_z$ is the momentum transferred in the scattering process\cite{cp_book,kaplan03}. High-resolution x-ray Compton scattering studies have revealed that the orbital involved in the lithium insertion/extraction process in Li$_x$Mn$_2$O$_4$ is mainly the oxygen $2p$ orbital\cite{Suzuki}. Although the oxygen $2p$ orbitals thus dominate the redox process, magnetic properties of the material are controlled by manganese $3d$ orbitals. Here, we use magnetic Compton scattering (MCS) to explore how the manganese $3d$ states evolve with Li intercalation and how their magnetism drives performance of the battery. 

In an MCS experiment, one measures the magnetic Compton profile (MCP)\cite{cp_book}, $J_{mag}$, which can be expressed in terms of a double integral of the spin-dependent electron momentum density, $\rho_{mag}({\mathbf p})$, as
\begin{equation}
J_{mag}(p_z)= \int \int  \rho_{mag}({\mathbf p}) dp_x dp_y.
\label{eq2}
\end{equation}
Here, $\rho_{mag}({\mathbf p}) = \rho_{\uparrow}({\mathbf p}) - \rho_{\downarrow}({\mathbf p})$, where $\rho_{\uparrow}({\mathbf p})$ and $\rho_{\downarrow}(\mathbf p)$ are the momentum densities of up-spin (majority) and down-spin (minority) electrons, respectively. The area under the magnetic Compton profile $J_{mag}(p_z)$ yields the total magnetic spin moment. Therefore, MCS experiments require a strong magnetic field and a net total magnetic moment in the sample. In this way, MCS allows access to magnetic electrons in materials and its potential in this regard was recognized quite early in the field\cite{platzman65,sakai76}. The technique has proven especially successful in extracting occupation numbers of 3$d$ Mn orbitals in bilayer manganites\cite{koizumi,li} and recent MCS studies have revealed fine details of the magnetic orbitals in a number of materials\cite{Duffy,Kamali,Gillet}.

\begin{figure}[t!]
\centering
\includegraphics[scale=0.30]{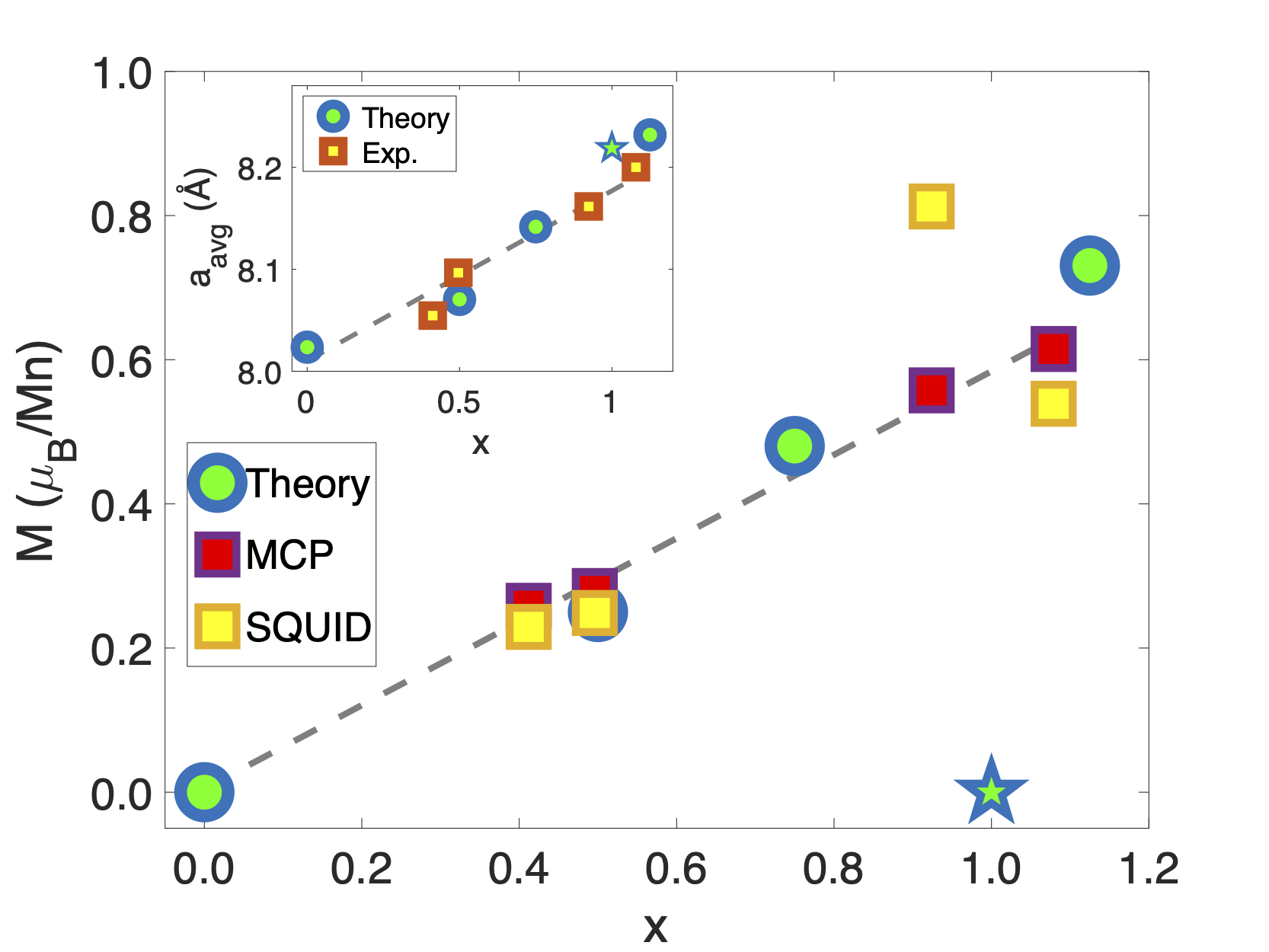}
\label{fig_anomaly}
\caption{FIM phase in Li$_x$Mn$_2$O$_4$. Magnetic moments for various lithium concentrations $x$ obtained via magnetic Compton scattering (red squares) and SQUID (yellow squares) measurements are compared with the corresponding theoretical predictions (green circles). Disappearance of the total magnetic moment at $x=1$ is highlighted by the star symbol. The inset shows how the experimental and theoretical values of the average lattice constant vary with $x$ and approximately follow a linear behavior consistent with Vegard's law.}
\end{figure}

\begin{figure}
\centering
\hspace*{-7mm}
\includegraphics[scale=0.3]{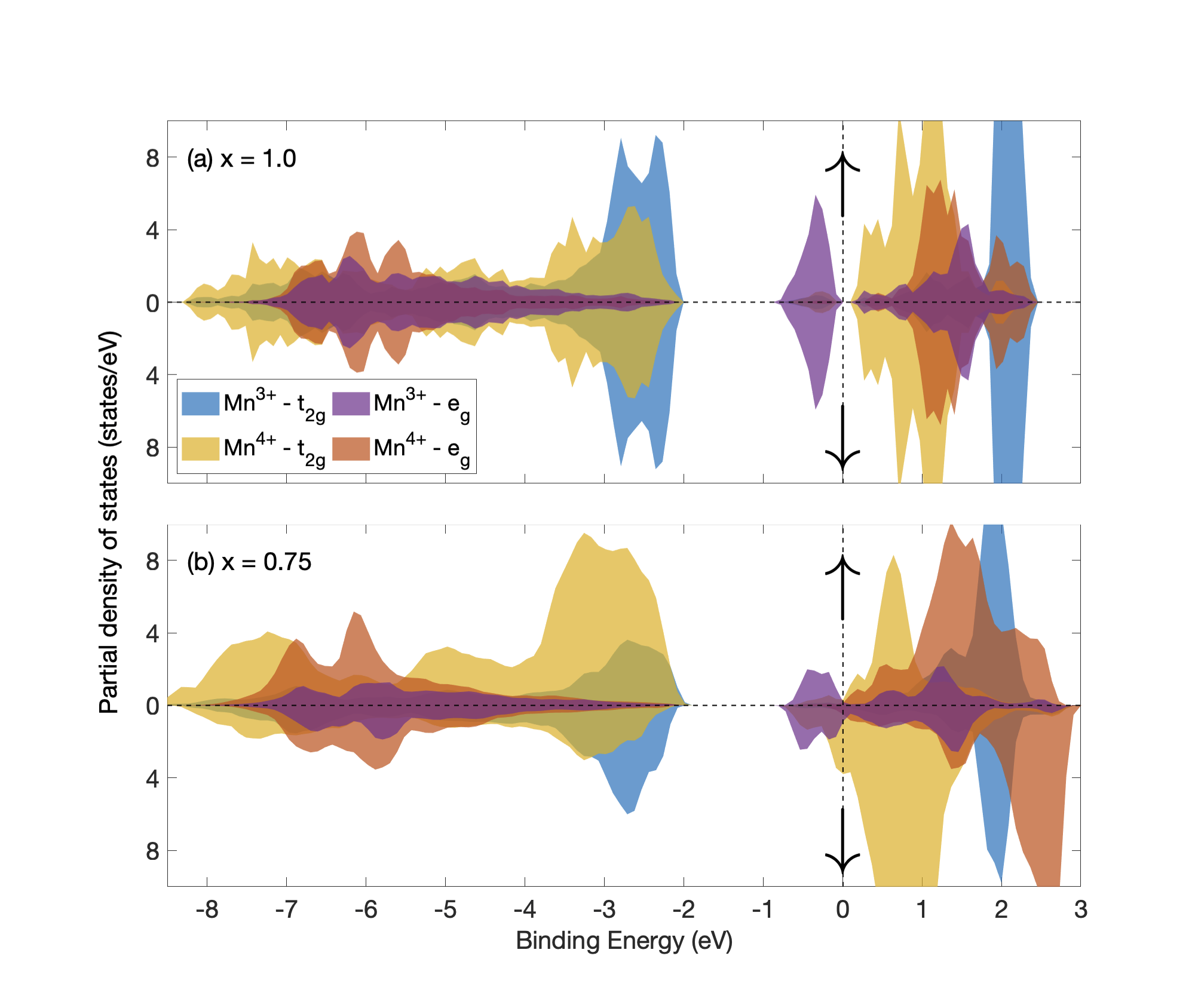}
\label{fig_dos}
\vspace*{-5mm}
\caption{Computed spin-dependent partial-density-of-states (PDOS) associated with the e$_g$ and t$_{2g}$ orbitals of Mn$^{3+}$ and Mn$^{4+}$ ions in Li$_x$Mn$_2$O$_4$ for Li concentrations (a) $x=1.0$ and (b) $x=0.75$. See panel (a) for the meaning of lines of various colors. Vertical dashed-line marks the Fermi energy (E$_F$). Up and down arrows indicate the contributions of spin up and down PDOS, respectively.}
\end{figure}

\begin{figure*}
\centering
\hspace*{-5mm}
\includegraphics[scale=0.45]{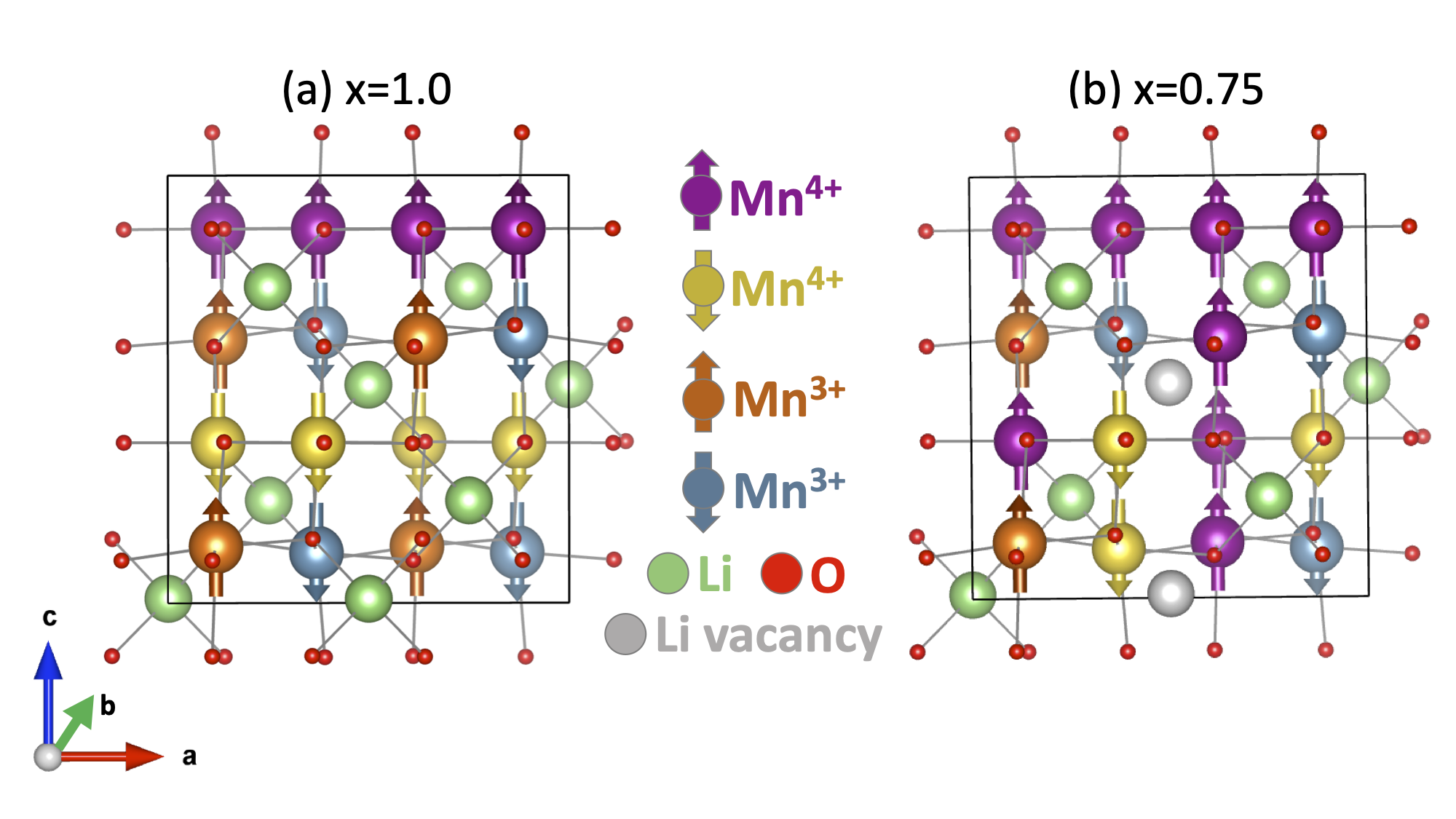}
\label{fig:3}
\vspace*{-5mm}
\caption{Schematics of magnetic configurations of Li$_x$Mn$_2$O$_4$ for lithium concentrations $x=1$ and $x=0.75$. Charge-ordering is prominent for the $x=1$ phase with an overall AFM configuration (a). As soon as the Li ions are removed, see (b), charge-ordering becomes unstable, and results in partial ordering, spin-flipping, and reduction of the average local magnetic moment per Mn atom. This induces ferrimagnetism at $x=0.75$, which is seen also in the PDOS of Fig.~2(b). }
\vspace*{10mm}
\end{figure*}

\section{MATERIALS AND METHODS}

\subsection{Magnetic Compton experiments}
The MCS experiments were carried out at $10$ K on beamline BL08W at SPring-8, Japan\cite{Sakurai1998,Kakutani2003}. Elliptically polarized x-rays emitted from an elliptical multipole wiggler were monochromatized to $183$ keV by a bent Si 620 crystal. Energy spectra of Compton-scattered x-rays from the sample at a scattering angle of $178^{\circ}$ were measured using a ten-segmented Ge solid-state detector with external magnetic field of 25 kOe. The estimated momentum resolution is 0.50 a.u. full-width-at-half-maximum. The spherically averaged profile $J_{mag}(p)$ was extracted from the difference between two spectra taken under the same experimental conditions with alternating directions of magnetization of the sample, aligned by an external magnetic field\cite{cp_book}. The observed spectra were corrected for the energy-dependent scattering cross section, efficiency of the detector, and absorption of the sample.

Polycrystalline samples of Li$_x$Mn$_2$O$_4$ ($x= 0.41$, $0.50$, $0.92$ and $1.08$) were prepared by chemical lithium extraction following the method reported previously\cite{Suzuki}. The compositions were determined by inductively coupled plasma (ICP) measurements. X-ray powder diffraction analyses confirmed spinel phases and the increase of the lattice constant with increasing $x$ over the range $x=0-1$ as observed previously\cite{Suzuki}. Total magnetic moments were obtained from SQUID magnetometer (MPMS5-SW, Quantum Design, Inc.) measurements.

\subsection{First-principles calculations}
First-principles calculations were performed using the pseudopotential projector augmented-plane-wave method\cite{paw} as implemented in the Vienna Ab-Initio Simulation Package (VASP)\cite{vasp1,vasp2}, with a kinetic energy cutoff of $600$ eV for the plane-wave basis set. Computations were carried out using both the Generalized Gradient Approximation (GGA)\cite{PBE,BMJ} and the recently constructed strongly-constrained-and-appropriately-normed (SCAN) meta-GGA\cite{SCAN} exchange-correlation functional. A 4$\times$4$\times$4 $\Gamma$-centered $k$-point mesh was used to sample the Brillouin zone of a primitive spinel unit cell (containing eight formula units with 32 oxygen atoms). Equilibrium positions of all atoms were calculated via structural optimization, where the internal degrees of freedom, along with the shape and volume of the unit cell, were allowed to vary until the residual forces per atom were less than 0.005 eV/\AA. We obtained spin momentum densities, $\rho_{mag}({\mathbf p})$, and the spherical Compton profiles $J_{mag}(p)$ of the valence electrons using Kohn-Sham orbitals following the method of Makkonen {\em et al.}\cite{Makkonen}. This scheme has been recently used to study the Compton profile of lithium iron phosphate (LiFePO$_4$)\cite{Hasnain}, which is as an exemplar cathode battery material\cite{matteo2019, Wanli_JACS}.

\section{RESULTS AND DISCUSSIONS}

\subsection{Ferrimagnetic (FIM) phase}
Figure~1 highlights the evolution of the magnetic moment in Li$_x$Mn$_2$O$_4$ for Li concentrations varying from zero to around 1.1 per formula unit. The experimental Compton magnetic moment is seen to follow a linear behavior as shown by the gray dashed line. The AFM state (zero average moment) is realized in the end compounds MnO$_2$ and LiMn$_2$O$_4$, but the system assumes an FIM state with a net non-zero moment at other compositions, including the over-discharged regime of $x>1$. The moments obtained via SQUID measurements are in good accord with those obtained from MCS experiments.  Some differences are to be expected since MCS measures only the spin moment\cite{platzman65} while the SQUID couples to the total moment, which includes the orbital component of the magnetic moment. Notably, anomalous behavior of the magnetic moment near the Verwey transition has been reported previously \cite{li2007}. Our theoretical moments (green circles) are in reasonable accord with the corresponding experimental results (red and yellow squares). Some excursion from linearity in theoretical moments could be due to the small size of the simulation cell used in computations. In fact, Korringa-Kohn-Rostoker-coherent-potential-approximation (KKR-CPA)\cite{KKRCPA2001, KKRCPA1999} calculations show that the spin moment on Mn atoms increases linearly with increasing $x$ if the system is described by a spin-glass-like behavior with randomly oriented Mn moments\cite{Suzuki}. Moreover, the strengthening of Mn moments with Li insertion is consistent with muon-spin-rotation experiments\cite{mukai}. However, the random-alloy model\cite{Suzuki} underlying the KKR-CPA scheme cannot be expected to capture the AFM charge-ordered state at $x=1$.

\subsection{Charge, spin and magnetic configuration}
The SCAN functional used in this study successfully localizes Mn $3d$ electrons by filling  $3d_{z^2}$ orbitals, and results in the coexistence of Mn$^{3+}$ and Mn$^{4+}$ in LiMn$_2$O$_4$.  Figure~2, which presents our spin-dependent partial-density-of-states (PDOS) in Li$_x$Mn$_2$O$_4$, shows that $e_g$ orbitals of some Mn atoms split upon lithium intercalation. The Mn $3d_{z^2}$ levels move to lower energies and become filled, and eventually open a band gap of about 0.1 eV at $x=1$. For $0<x<1$, the system is metallic due to the partial occupation of the Mn $3d_{z^2}$ bands. At $x=1$, the ground-state is found to be an AFM charge-ordered state with alternating AFM Mn$^{3+}$ layers and FM Mn$^{4+}$ layers with spins along the (001) direction as illustrated in Fig.~3(a). FIM states appear as soon as Li atoms are removed from or added to the unit cell.  A stable FIM configuration for $x=0.75$ is shown in Fig.~3(b). In this phase, the spin of the $t_{2g}$ electrons is not compensated according to the PDOS shown in Fig.~2(b), and therefore, these electrons produce a net magnetic moment. Interestingly, we found a similar magnetic PDOS for the  over-lithiated phase $x=1.125$ where the Mn $t_{2g}$ state is partially compensated, and hence, it is mainly responsible for the magnetic moment as shown in Supplementary Fig.~S1(a). In general, SCAN is able to describe complex configurations of competing stripe and magnetic phases in the cuprates \cite{SCAN_Yubo} and it also produces a small distortion from the $Fd3m$ cubic symmetry of Li$_x$Mn$_2$O$_4$ at $x=0.5$ (See Figs.~S1(b) and S3(b) in Supplemental Material\cite{Supp}), which  is consistent with the recent experimental results of Bianchini {\em et al.}\cite{Bianchini} In sharp contrast to the preceding SCAN-based results, the GGA functional\cite{Grechnev} fails to produce the insulating AFM state of LiMn$_2$O$_4$, and the associated charge disproportionation on Mn atoms that drives local Jahn-Teller distortions\cite{liu2017} of the MnO$_6$ octahedra. Notably, the theoretical lattice parameters obtained via SCAN are in good agreement with the corresponding experimental values as shown in the inset of Fig.~1.

\begin{figure}
\centering
\includegraphics[scale=0.70]{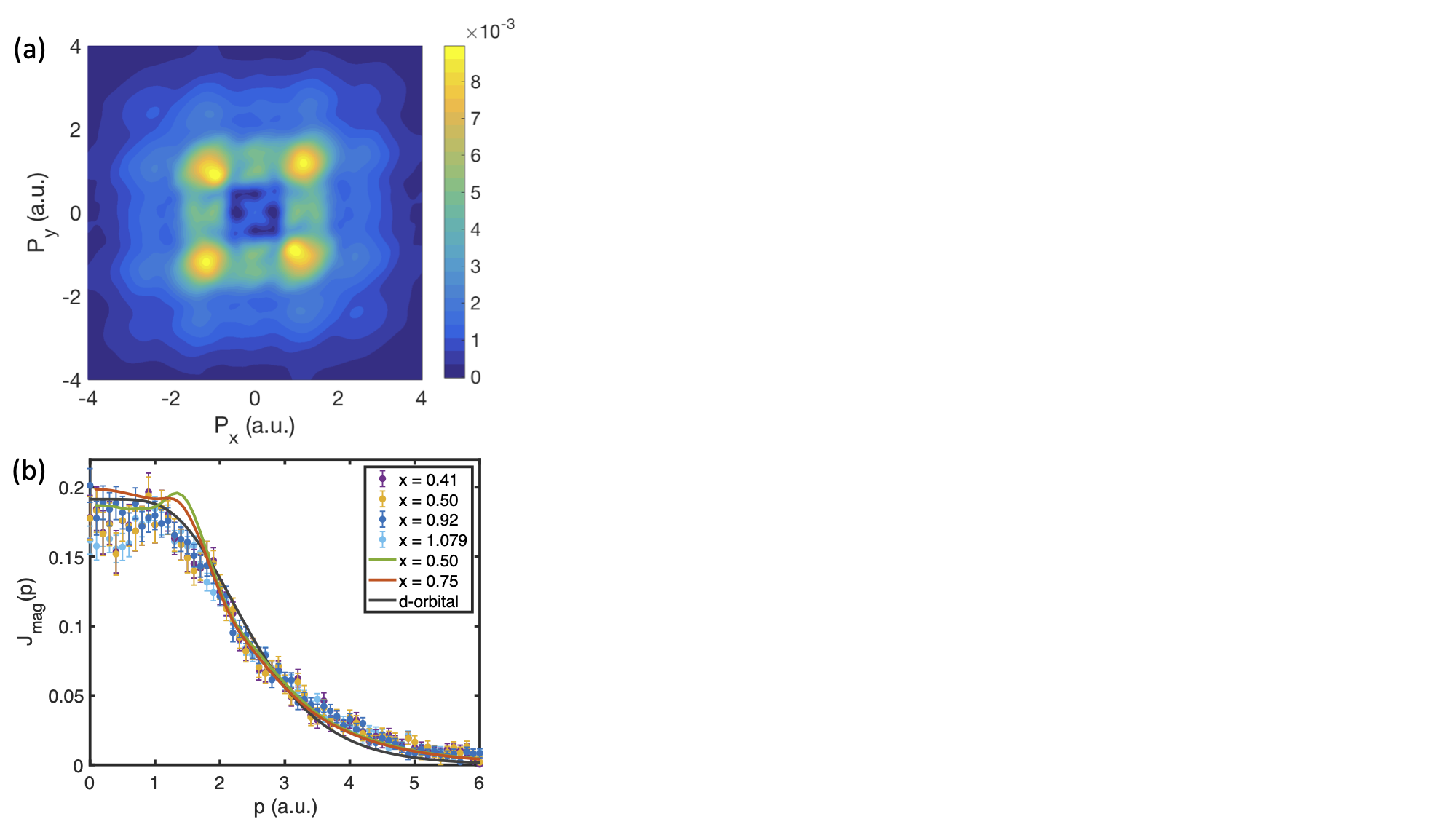}
\label{fig:4}
\caption{Compton profiles and visualization of the magnetic orbitals. (a) Theoretical 2D momentum density map for projection along (001) for $x=0.75$. (b) Experimental (scatter plots with error bars) and theoretical (solid lines) spherically-averaged magnetic Compton profiles. Grey solid-line represents atomic $d$-orbital Compton profile which is solved analytically using normalized Slater-type orbitals. All the profiles are normalized to one.}
\vspace*{10mm}
\end{figure}

\subsection{Magnetic orbitals and Compton profiles.}
Electronic states in momentum space preserve their individual angular dependencies, which facilitates the detection of particular states\cite{Harthoorn}. Thus, the projection of the theoretical 3D spin-polarized electron momentum density $\rho_{mag}({\mathbf p})$ along the $(001)$ direction at $x=0.75$ given in Fig.~4(a) determines the nature of the electronic states producing the FIM solution. Clearly, the angular dependence of $\rho_{mag}({\mathbf p})$ is dominated by the $t_{2g}$ character of the $3d$ Mn electrons \cite{TChiba}.  This spin momentum result is consistent with our previous observation that the $t_{2g}$ electrons are not compensated according to Fig.~2(b) of the PDOS. Based on these robust arguments, one can conclude that the FIM orbitals have mostly $t_{2g}$ character. Figure~4(b) presents the measured and calculated spherically averaged MCPs for various lithium concentrations, $x$. For all values of $x$, the shapes of the MCPs are quite similar, the main difference being the area under the profile, which gives the total spin magnetic moment. As we noted already, the spin magnetic moment follows a linear behavior for all lithium concentrations with the exception of the $x=1$ compound where the moment vanishes. The agreement between theory and experiment in Fig.~4(b) is considered very good when we keep in mind the delicate nature of the MCP. The shape of the experimental $J_{mag}(p)$ is also similar to that of the Compton profile from a manganese $3d$ atomic orbital\cite{bb_dorbitalcp} since the manganese $t_{2g}$ orbitals undergo little hybridization with the $2p$ orbital of oxygen\cite{Suzuki}. Figure~5 shows a comparison of the directional Compton profiles between the KKR-CPA and SCAN calculations. As mentioned above, our KKR-CPA calculations describe a spin-glass-like behavior with randomly oriented Mn moments embedded in a perfect cubic spinel structure. Therefore, the agreement between the two methods indicates that SCAN can successfully capture the almost cubic FIM spin-glass phase. Thus, the Jahn-Teller distortions obtained within SCAN are strongly suppressed if the electrons occupy the FIM orbitals visualized in Fig.~4(a). As shown by our magnetic Compton scattering experiments, the application of an external field promoting the FIM phase and the associated orbitals can serve as a basis for preventing Jahn-Teller distortions. 

\begin{figure}[t!]
\centering
\hspace*{-5mm}
\includegraphics[scale=0.42]{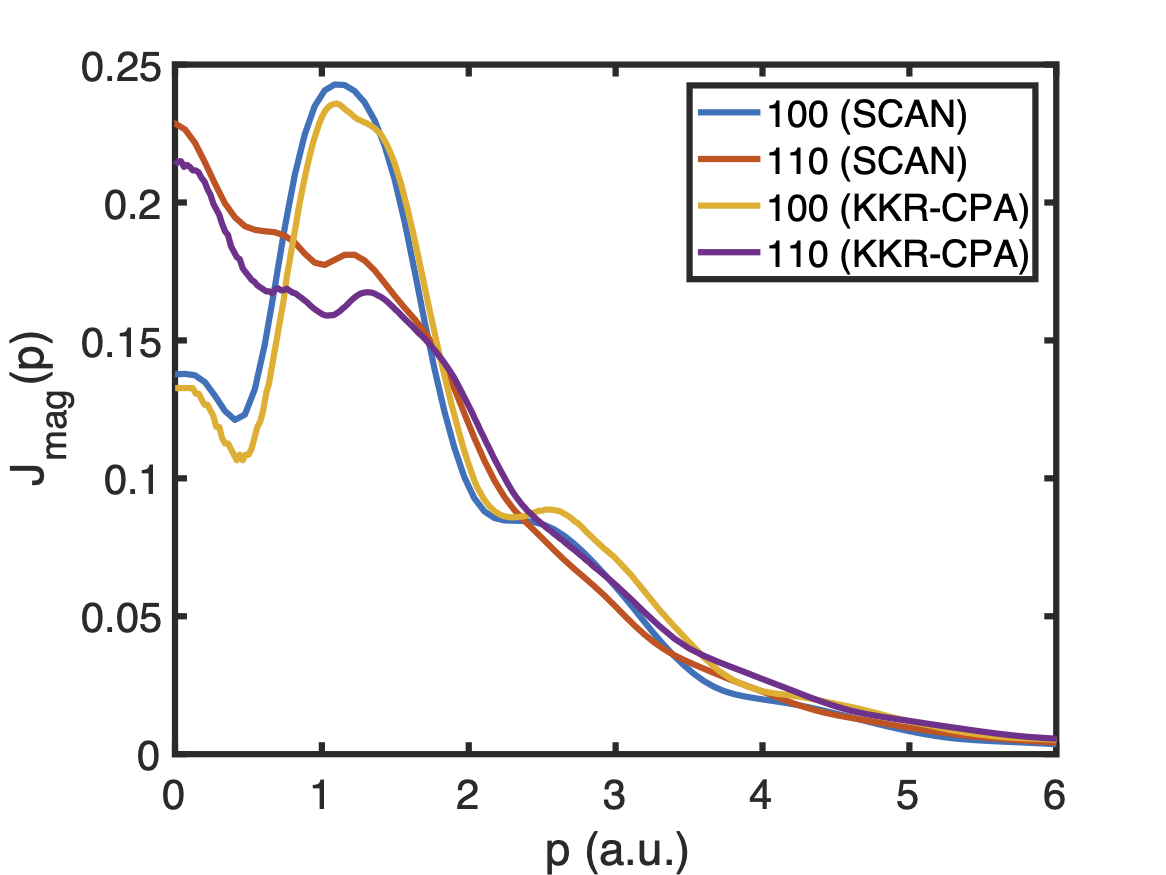}
\label{fig:5}
% \vspace*{-20mm}
\caption{Directional Magnetic Compton profiles along the $(100)$ and $(110)$ directions. The low-momentum contributions mainly come from the $(110)$ directions, which contribute significantly to the spherical average in Fig.~4. The SCAN results show good agreement with the KKR-CPA results for the FIM phase.}
\vspace*{10mm}
\end{figure}

\section{Conclusions}
We have shown that a fundamental understanding of the role of magnetic electrons in Li-ion battery materials can be obtained via magnetic Compton scattering experiments and first-principles calculations.  Specifically, the magnetic Compton profiles of Li-Mn-O system at different lithium concentrations have been analyzed to unfold the underlying magnetic transitions in terms of the spin-moment contribution of the Mn $3d$ orbitals. The calculated total spin moments are in good agreement with the corresponding experimental values. The present analysis demonstrates that the Mn $3d$ magnetic electrons have mainly $t_{2g}$ symmetry, and therefore, these electrons prevent Jahn-Teller distortions promoted by the $e_g$ character.  Our study explains the intimate connection between the charge, spin and lattice degrees of freedom and their role in the bonding of the MnO$_6$ octahedra\cite{Whittingham} in the spinel battery materials. Although our present analysis is based on spherically-averaged experimental MCPs, it will be interesting to consider directional MCPs from single-crystal samples to gain further insight into the nature of the connection between the octahedral bonds and magnetism. Electron transfer during the lithiation and delithiation processes is shown to involve octahedral units whose distortion can be affected by temperature and magnetic field. In this way, the nature of the octahedral bonds and magnetism are connected with the electrochemical performance of the battery materials. 
\\\\

\section*{ACKNOWLEDGMENTS}
We thank Dr.~M.~Itou for technical support in connection with magnetic Compton scattering experiments. K.S. was supported by a Grant-in-Aid for Young Scientists (B) (No.~15K17873) from the Ministry of Education, Culture, Sports, Science, and Technology (MEXT), Japan, and the work at JASRI was partially supported by the Japan Science and Technology Agency. Compton scattering experiments were performed with the approval of JASRI (Proposals Nos.~2014B1335 and 2015B1171). SQUID measurement was performed with the approval of Gunma University-Industry Collaboration and Intellectual Property Strategy Center (Proposal 2017). The work at Northeastern University was supported by the U.S. Department of Energy, Office of Science, Basic Energy Sciences Grant No.~DE-FG02-07ER46352, and benefited from Northeastern University’s Advanced Scientific Computation Center (ASCC), and the allocation of time at the NERSC supercomputing center through DOE Grant No.~DE-AC02-05CH11231. S. K. was supported by the Polish National Science Center (NCN) under Grant No.~DEC-2011/02/A/ST3/00124. \\

H.H. and K.S. contributed equally to this work.

\bibliography{main.bib}

\beginsupplement

\begin{figure*}
\centering
% \hspace*{-5mm}
\includegraphics[scale=0.6]{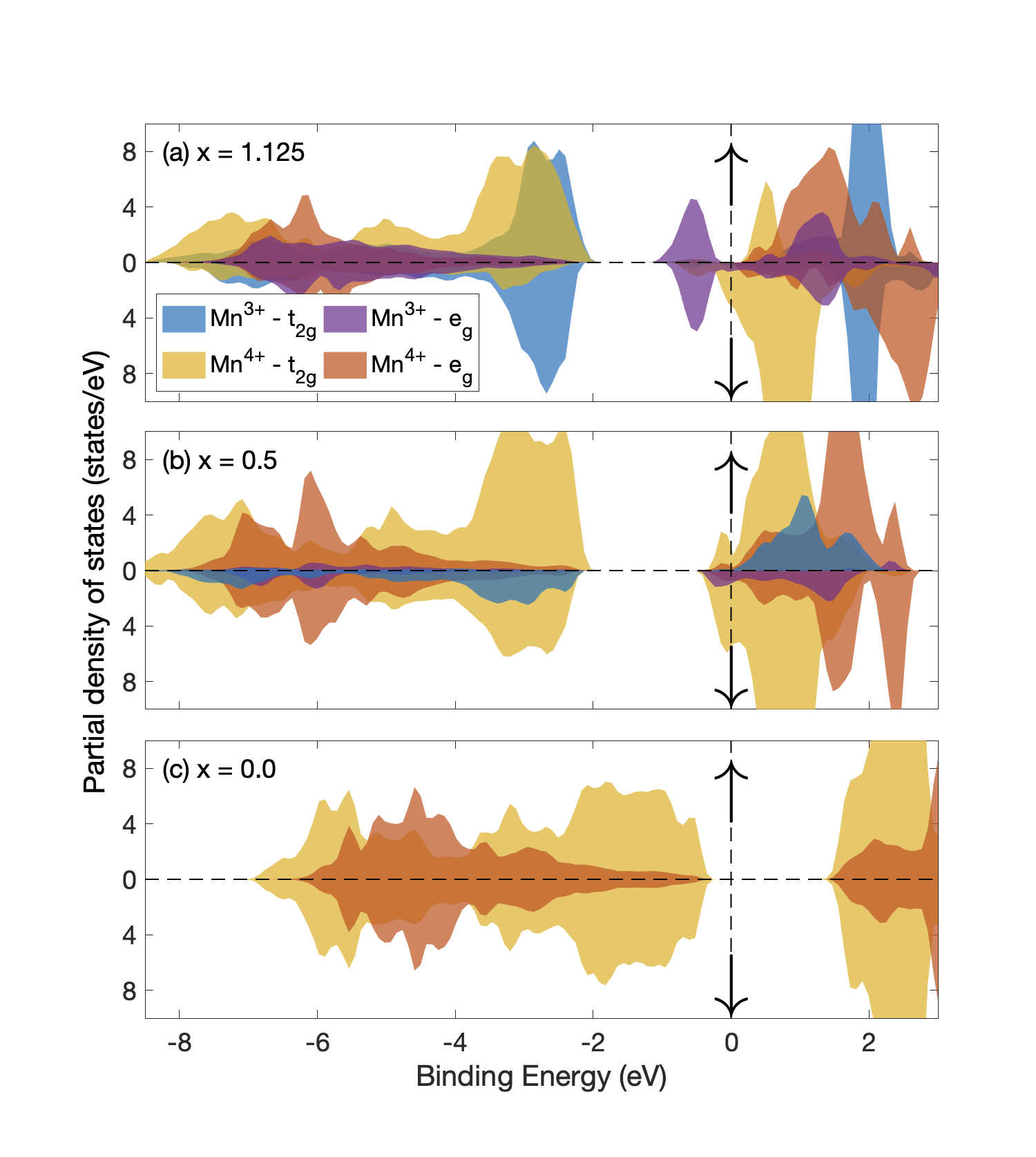}
\label{fig_S1}
\caption{Computed spin-dependent partial-density-of-states (PDOS) associated with the e$_g$ and t$_{2g}$ orbitals of Mn$^{3+}$ and Mn$^{4+}$ ions in Li$_x$Mn$_2$O$_4$ for Li concentrations (a) $x=1.125$, (b) $x=0.5$ and (c) $x=0.0$. See panel (a) for meaning of lines of various colors. Vertical dashed line marks the Fermi energy (E$_F$). Up and down arrows indicate the contributions of spin up and down PDOS respectively. SCAN results yield insulating states for $x=0.0$ and $1.0$ (Fig. 2a), and metallic states for $x=1.125, 0.75$ (Fig. 2b), and $0.5$.}
\vspace*{25mm}
\end{figure*}

\begin{figure*}
\centering
% \hspace*{-5mm}
\includegraphics[scale=0.7]{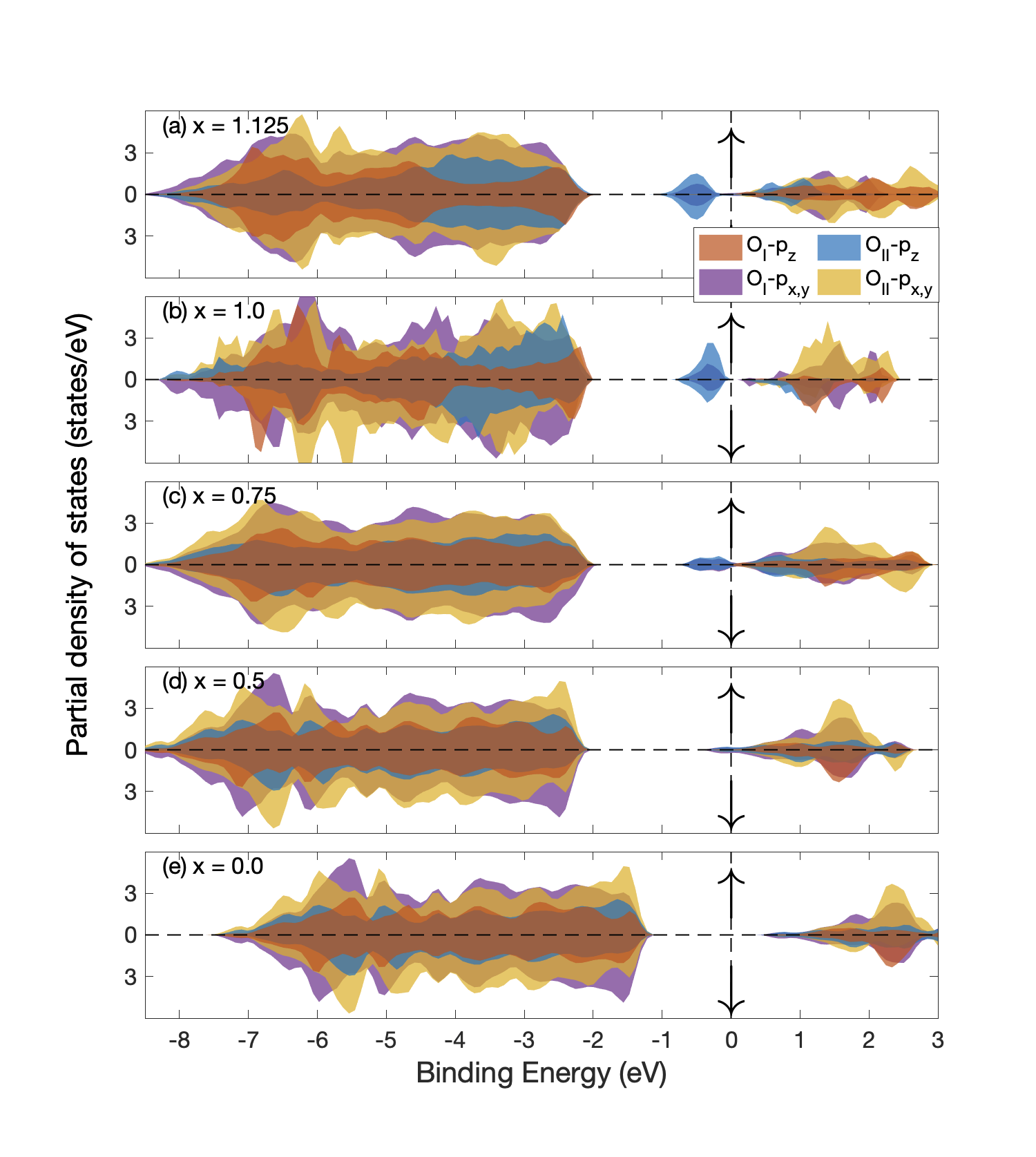}
\label{fig_S2}
% \vspace*{-15mm}
% \renewcommand{\figurename}{Fig.S2}
\caption{Computed projected density-of-states of O-2p states. O$_I$ and O$_{II}$ atoms indicate the oxygen atoms from the Mn$^{4+}$ and Mn$^{3+}$ layers respectively for the AFM configuration in $x=1$.}
\vspace*{15mm}
\end{figure*}

\begin{figure*}
\centering
\includegraphics[scale=0.85]{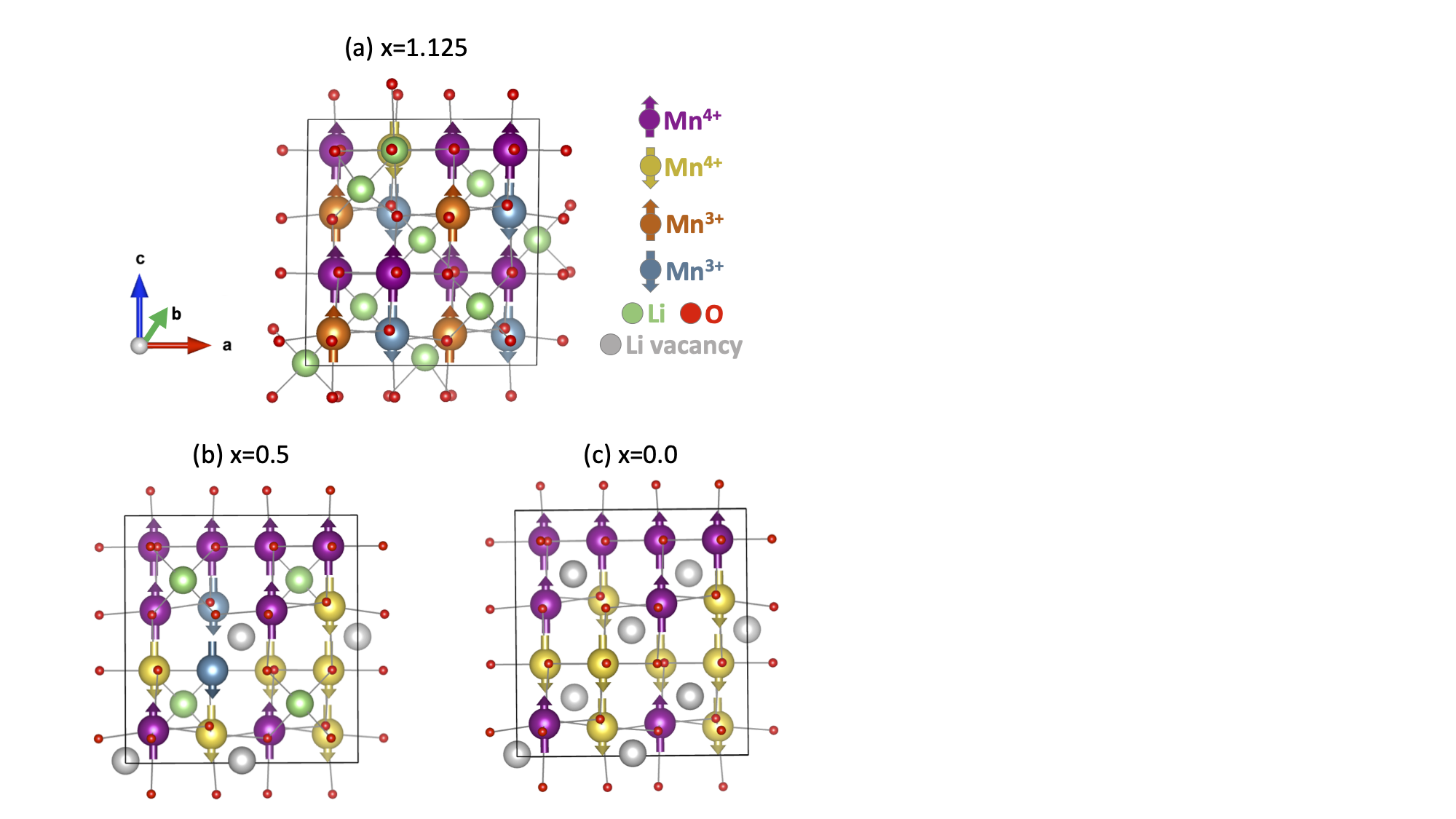}
\label{fig_S3}
\caption{Schematics of magnetic configurations of Li$_x$Mn$_2$O$_4$ for lithium concentrations, (a) $x=1.125$, (b) $x=0.5$ and (c) $x=0.0$. All the structures initially assume a cubic symmetry with space group Fd3m (227) where Li, Mn and O atoms occupy the Wyckoff positions: $8a$, $16d$ and $32e$ respectively. For over-lithiated case, the extra Li goes to the octahedral position $16c$ since all the tetrahedral positions $8a$ are occupied. DFT optimization shows a breaking of the cubic Fd3m symmetry during structural relaxation.}
\vspace*{110mm}
\end{figure*}

\begin{figure*}
\centering
% \vspace*{-15mm}
\hspace*{15mm}
\includegraphics[scale=0.8]{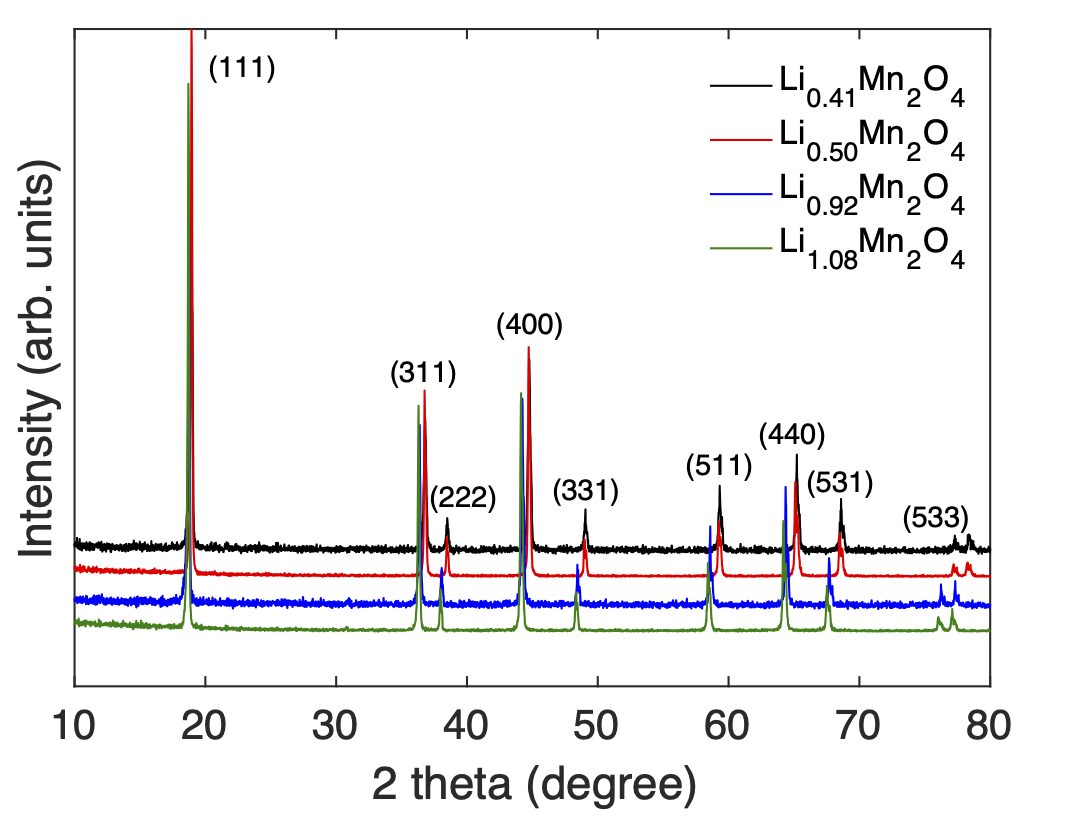}
\label{fig_S4}
% \vspace*{-15mm}
\caption{XRD pattern for Li$_x$Mn$_2$O$_4$ (x=0.41, 0.50, 0.92 and 1.08)}
% \vspace*{-50mm}
\end{figure*}

\begin{table*}
\begin{center}
\begin{tabular}{ |c|c| } 
\hline
Li concentration (x) & Lattice constant ($\angstrom$)\\
 \hline 
0.41 & 8.0546\\
0.5 & 8.0966\\
0.92 & 8.1616\\
1.08 & 8.2\\
 \hline 
\end{tabular}
\caption {Experimental lattice constant obtained from XRD analysis.}
\end{center}
\vspace*{-7mm}
\end{table*}

\end{document}